# GP-Tree: An in-memory spatial index combining adaptive grid cells with a prefix tree for efficient spatial querying

Xiangyang Yang [a], Xuefeng Guan [b, *], Lanxue Dang [a], Yi Xie [a], Qingyang Xu [b], Huayi Wu [b], Jiayao Wang [c, d]

[a] *School of Computer and Information Engineering, Henan University, Kaifeng, China;* [b] *State Key Laboratory of Information Engineering in Surveying, Mapping and Remote Sensing, Wuhan University, Wuhan, China;* [c] *State Key Laboratory of Spatial Datum, College of Remote Sensing and Geoinformatics Engineering，Faculty of Geographical Science and Engineering, Henan University, Zhengzhou, China;* [d] *Henan Industrial Technology Academy of Spatio-Temporal Big Data, Henan University, Zhengzhou, China*

**Abstract**

Efficient spatial indexing is crucial for processing large-scale spatial data. Traditional spatial indexes, such as STR-Tree and Quad-Tree, organize spatial objects based on coarse approximations, such as their minimum bounding rectangles (MBRs). However, this coarse representation is inadequate for complex spatial objects (e.g., district boundaries and trajectories), limiting filtering accuracy and query performance of spatial indexes. To address these limitations, we propose GP-Tree, a fine-grained spatial index that organizes approximated grid cells of spatial objects into a prefix tree structure. GP-Tree enhances filtering ability by replacing coarse MBRs with fine-grained cell-based approximations of spatial objects. The prefix tree structure optimizes data organization and query efficiency by leveraging the shared prefixes in the hierarchical grid cell encodings between parent and child cells. Additionally, we introduce optimization strategies, including tree pruning and node optimization, to reduce search paths and memory consumption, further enhancing GP-Tree's performance. Finally, we implement a variety of spatial query operations on GP-Tree, including range queries, ε-distance queries, and k-nearest neighbor queries. Extensive experiments on real-world datasets demonstrate that GP-Tree significantly outperforms traditional spatial indexes, achieving up to an order-of-magnitude improvement in query efficiency.

* Corresponding author. Email address: guanxuefeng@whu.edu.cn.

## 1. Introduction

The rapid growth of spatial data, driven by advancements in sensors, mobile devices, and satellite systems, has resulted in the generation of vast spatial data. For example, NASA's EOSDIS archive is expected to reach nearly 250PB by 2025, while Planet Labs gathers more than 15TB of satellite data daily (Shang and Eldawy 2023). In 2025 Q2, Uber had 180 million monthly active consumers worldwide, generating approximately 36 million trips per day on average (Kumar 2025). These large-scale spatial datasets are invaluable for various applications, such as suspected infected crowds detection (He *et al.* 2020) and event detection (George *et al.* 2021). However, efficiently processing such massive datasets presents significant challenges, primarily due to their multi-dimensional structure and the high computational demands involved.

Spatial indexing technology, which organizes spatial objects in a structured manner, is an effective approach for improving the efficiency of processing big spatial data. Over the past few decades, various spatial indexes have been proposed, broadly categorized into two types: single-entry spatial indexes and multi-entry spatial indexes (Schoemans *et al.* 2024).

Single-entry spatial indexes, such as the R-Tree family (Leutenegger *et al.* 1997, Vu and Eldawy 2020) and Quad-Tree (Hjaltason and Samet 1999), represent each spatial object using one index entry, typically a minimum bounding rectangle (MBR) for each spatial object. However, an MBR provides only a coarse approximation, especially for irregularly shaped polygons or trajectories, leading to large empty areas that do not contain any actual spatial objects (Sidlauskas *et al.* 2018), as illustrated in Figure 1(a). This coarse representation limits the filtering effectiveness and reduces the overall query performance of single-entry spatial indexes.

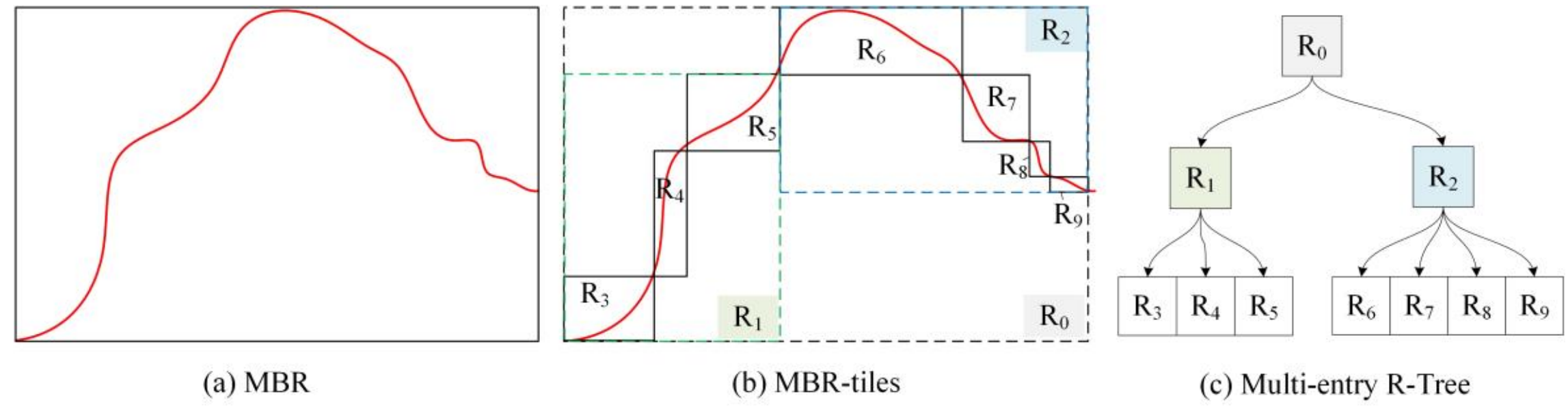

(a) MBR (b) MBR-tiles (c) Multi-entry R-Tree

Figure 1. Examples of single-entry and multi-entry spatial indexes.

Multi-entry spatial indexes represent spatial objects with finer approximations, such as grid cells or multiple MBRs, and assign multiple index entries to each object to enhance retrieval efficiency. For example, ACT (Kipf *et al.* 2020) splits each spatial object into a set of grid cells and indexes them into a query-efficient radix tree. MSP-GiST (Schoemans *et al.* 2024) represents a trajectory using multiple MBRs as shown in Figure 1(b), and supports multi-entry versions of traditional spatial index structures, e.g., R-Tree and Quad-Tree. Figure 1(c) shows a multi-entry R-Tree that indexes multiple MBRs in Figure 1(b). Multi-entry spatial indexes can improve filtering effectiveness and query performance by providing a finer-grained representation of spatial objects.

Despite significant advances in multi-entry spatial indexes, several limitations remain. First, as the number of index entries increases, maintaining efficient query processing becomes more challenging. Existing approaches primarily improve filtering effectiveness by representing each spatial object using multiple approximation units, such as multiple MBRs. However, these approaches typically continue to organize the approximations using conventional spatial index structures, such as R-Tree and Quad-Tree. Consequently, query processing still relies on geometric operations (e.g., MBR intersection) to determine node positions during index traversal, which may incur higher computational overhead as the number of approximation units increases.

Second, multi-entry indexes suffer from limited scalability. Mapping each object to multiple entries reduces the index's flexibility due to the tight coupling between object representations and query-specific processing schemes. For instance, ACT(Kipf *et al.* 2018) only supports range queries on points, limiting its applicability to a narrow range of spatial object types and query operations.

To address these limitations, we propose GP-Tree, a novel multi-entry spatial index that integrates fine-grained, grid-based approximations of spatial objects with a query-efficient prefix tree. First, GP-Tree adaptively divides spatial objects into multi-type, multi-resolution grid cells through a Discrete Global Grid System(Sahr *et al.* 2003). Unlike high-dimensional MBR-based representations, these grid cells can be encoded into one-dimensional identifiers via space-filling curves. These encodings naturally exhibit shared prefixes among parent and child cells. Based on this property, GP-Tree then organizes these grid cells into a prefix tree. Unlike traditional index structures (e.g., R-Tree and Quad-Tree), which perform coordinate-based comparisons during queries, the prefix tree enables more efficient querying through fast prefix matching on hierarchical grid encodings. Additionally, GP-Tree incorporates optimization strategies, including tree pruning and node optimization, to shorten search paths and reduce memory consumption, further improving GP-Tree's efficiency. GP-Tree also integrates Grid-AM in GridMesa(Yang *et al.* 2024) to improve geometric operations. Finally, GP-Tree supports efficient range, distance, and kNN queries across Points, Linestrings and Polygons.

The contributions of this paper are as follows:

(1) We propose GP-Tree, which combines the advantages of fine-grained grid-based approximations and the query-efficient structure of prefix tree.

(2) GP-Tree is equipped with optimization strategies, including tree pruning and node optimization, to shorten search paths and reduce the memory cost, further improving its performance.

(3) GP-Tree can flexibly and efficiently support a wide range of spatial queries (e.g., range, distance, and kNN queries) across different spatial object types.

(4) Extensive experiments on real-world datasets demonstrate the GP-Tree outperforms baselines by up to an order of magnitude in query performance.

The rest of the paper is organized as follows. Section 2 reviews related work on spatial indexes. Section 3 provides the necessary background. Sections 4 and 5 introduce the structure and query operations of GP-Tree in detail, respectively. Section 6 provides a comprehensive evaluation of GP-Tree, and Section 7 concludes the paper.

## 2. Related work

Spatial indexes are categorized into single-entry and multi-entry spatial indexes (Schoemans *et al.* 2024).

**Single-entry spatial indexes.** Traditional single-entry spatial indexes, such as STR-Tree(Leutenegger *et al.* 1997) and KDB-Tree(Robinson 1981), aim to enhance the filtering efficiency of spatial queries by using MBRs as approximations of spatial objects. Hilbert R-tree (Kamel and Faloutsos, 1994) employs the Hilbert space-filling curve to impose a linear ordering on MBRs, thereby improving spatial locality and constructing a more compact R-tree with reduced node overlap. Popular spatial management systems, such as GeoSpark(Yu *et al.* 2015), Simba(Xie *et al.* 2016) and Apache Sedona(García-García *et al.* 2021), are all equipped with these indexes. Since MBRs provide only coarse approximations of spatial objects, MBR-based filtering results in a large number of mismatched candidates requiring subsequent time-consuming geometric refinement. To address the limitation, several improved single-entry indexes have been proposed. The clipped bounding box (CBB)(Sidlauskas *et al.* 2018) reduces dead space in MBRs by clipping away redundant corners using a few auxiliary points. It can be seamlessly integrated into existing R-tree variants to improve filtering effectiveness. GE-Tree(Shin *et al.* 2019) is a hybrid spatial index, in which grid cells are incorporated into the leaf level of a Quad-Tree to facilitate tree navigation and maintenance. Each grid cell maintains a pointer to the corresponding leaf node which contains spatial objects. MaMBo(Evagorou and Heinis 2021) uses uniform grid cells to index dead space in addition to an index of the spatial objects. It then uses the grid index to check if queries lead to dead space, enabling early termination of queries and reducing the number of leaf node accesses.

**Multi-entry spatial indexes** manage finer approximations of spatial objects to improve retrieval efficiency. D-Grid(Xu *et al.* 2016) is an in-memory dual space grid index for moving objects. Specifically, it indexes moving objects using grid structures in both location and velocity spaces. ACT(Kipf *et al.* 2018) approximates polygons as a set of grid cells and then stores the one-dimensional encodings of these cells into an in-memory radix tree to support efficient queries. CQG(Chen *et al.* 2023) combines a Quad-Tree and grid index to support complex spatial objects, where each node of the Quad-Tree is a grid cell rather than an MBR. The multi-entry generalized search tree (MGiST)(Schoemans *et al.* 2024) enables the decomposition of complex objects into multiple entries. It further extends traditional spatial indexing structures, such as R-tree, Quad-tree, and KD-tree, to support multi-entry searches, enhancing their filtering capability.

Table 1 provides a comparative summary of GP-Tree proposed in this paper against representative spatial indexes. Single-entry indexes, relying on a single-MBR representation, cannot overcome the query performance limitations caused by coarse approximations. While multi-entry indexes provide finer approximations, they struggle with scalability due to limited support for diverse spatial object types and query operations. In contrast, GP-Tree utilizes an adaptive grid–based approximation to offer finer object representation and efficiently supports range, distance, and kNN queries across Points, Linestrings, and Polygons. Furthermore, GP-Tree integrates Grid-AM in GridMesa (Yang *et al.* 2024) to improve geometric operations.

Table 1. Comparing GP-Tree against other spatial indexes.

| Type | Index | Approximation model | Query operation | Spatial object | Spatial library |
|---|---|---|---|---|---|
| Single-entry spatial indexes | STR-Tree (Leutenegger *et al.* 1997) | MBR | Range, distance, and kNN queries | Point, Polygon, Linestring | JTS |
| | CBB (Sidlauskas *et al.* 2018) | clipped MBR | Range query | Point, rectangle | JTS |
| | MaMBo (Evagorou and Heinis 2021) | MBR and cells | Range query | Point, rectangle | JTS |
| Multi-entry spatial indexes | ACT (Kipf *et al.* 2018) | grid cells | Range query | Point | JTS |
| | CQG (Chen *et al.* 2023) | grid cells | Range and kNN queries | Point, Polygon, Linestring | JTS |
| | MGiST (Schoemans *et al.* 2024) | MBR-tiles | Range and kNN queries | Linestring | JTS |
| | **GP-Tree** | **adaptive grid cells** | **Range, distance, and kNN queries** | **Point, Polygon, Linestring** | **Grid-AM** (Yang *et al.* 2024) |

## 3. Preliminaries and symbols

This section lists the frequently used symbols and their meanings and introduces the background on approximating spatial objects using grid cells.

### *3.1. Symbols*

For the purpose of reference, Table 2 lists the symbols and their meanings used frequently in this paper.

Table 2. Symbols and their meanings.

| symbols | description |
|---|---|
| $S$ | a spatial dataset |
| $s$ | a spatial object from $S$ |
| $MBR$ | minimum bounding rectangle |
| $SEG$ | the number of segments contained in each grid cell |
| $q$ | a spatial object for querying |
| $\varepsilon$ | $\varepsilon$ of $\varepsilon$-distance queries |
| $k$ | $k$ of kNN queries |

### *3.2. Grid-based approximation of spatial objects*

Studies have shown that grid-based approximations of spatial objects can improve spatial processing efficiency(Georgiadis and Mamoulis 2023, Yang *et al.* 2024). Grid-based approximations involve decomposing the Earth's surface using a hierarchical Quad-Tree structure. These decomposed grid cells are then enumerated by a space-filling curve (e.g., Hilbert or Z-order curves), which maps them into a one-dimensional data structure. In our implementation, we adopt the Z-order curve. Figure 2(a) illustrates how the cell encoding is computed using the bit-interleaving method of the Z-order curve. By leveraging the shared prefix property between parent and child cell encodings, spatial relationships between cells (e.g., contained and disjoint) can be efficiently determined using bitwise operations, as shown in Figure 2(b).

Figure 2(c) shows spatial objects and their approximated grid cells. Specifically, a point object is approximated as one grid cell; A non-point object (i.e., Linestring or Polygon) is adaptively approximated as multi-type grid cells, which are divided into interior grid cells (marked in blue) and boundary grid cells (marked in orange). In addition, each approximation cell has three attributes: "level", "id" and "isInterior" to identify itself and its relationship to the spatial object.

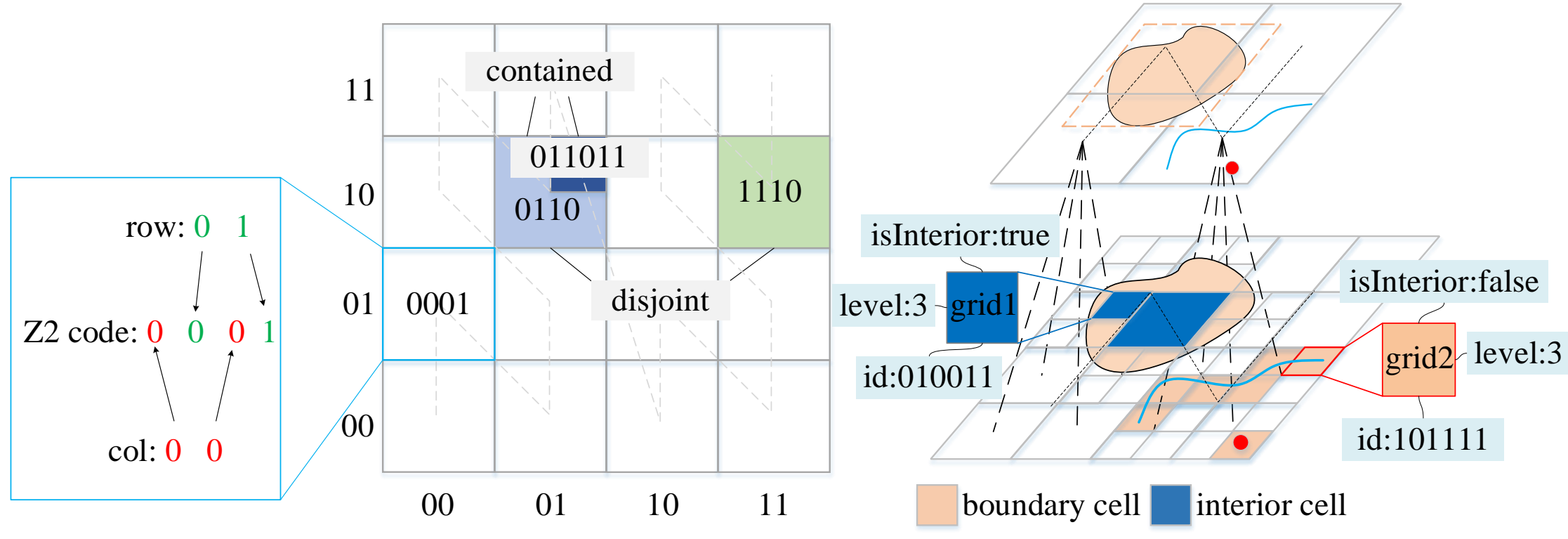

(a) Grid encoding in Z-ordering (b) Spatial relationships between grid cells (c) Grid-based approximations

Figure 2. Grid-based approximations for spatial objects.

## 4. The structure of GP-Tree

This section first introduces the basic structure of GP-Tree, then presents several optimization strategies to further enhance its performance, and finally analyzes its index construction complexity.

### *4.1. Basic GP-Tree*

GP-Tree is a hierarchical multi-entry spatial index that integrates adaptive grid cells of spatial objects with a prefix-tree indexing structure. In GP-Tree, each spatial object is approximated by a set of adaptive grid cells, providing a finer-grained representation than conventional single or multiple MBR approximations. These grid cells can be encoded into one-dimensional encodings, enabling spatial relationships among grid cells to be evaluated through encoding-based operations rather than costly geometric computations. Moreover, shared prefixes among hierarchical encodings allow grid cells to be compactly organized into a prefix tree and efficiently traversed through prefix matching. Consequently, GP-Tree improves both filtering effectiveness and query efficiency by combining fine-grained grid cells with an encoding-aware indexing structure.

Specifically, GP-Tree consists of two components: (i) a prefix tree that indexes grid cell encodings and stores spatial object IDs, and (ii) a lookup table that maps object IDs to their corresponding spatial geometries. Figure 3 illustrates the structure of GP-Tree.

The prefix tree uses grid cell encodings as keys, where each key each key corresponds to a path of a grid cell in the tree. GP-Tree consists of three types of nodes: the root node, intermediate nodes, and leaf nodes. The root node represents the global spatial scope and serves as the entry point of the index. Intermediate and leaf nodes represent shared prefixes of hierarchical cell encodings and maintain references to their child nodes. Figure 3(b) shows the basic GP-Tree structure constructed from the four polygons and their corresponding grid cells in Figure 3(a).

Each level of the tree consumes 2 bits of the cell encoding, meaning that each node may contain up to four child nodes corresponding to the four possible 2-bit values {0,1,2,3}. Consequently, a path from the root node to a given node uniquely determines the hierarchical encoding of a grid cell. For example, the red path in Figure 3(b) traverses the encoding sequence "10 → 10 → 00 → 11" forming the grid cell encoding “10100011”. This encoding corresponds to the interior grid cell of $O_1$ shown in Figure 3(a).

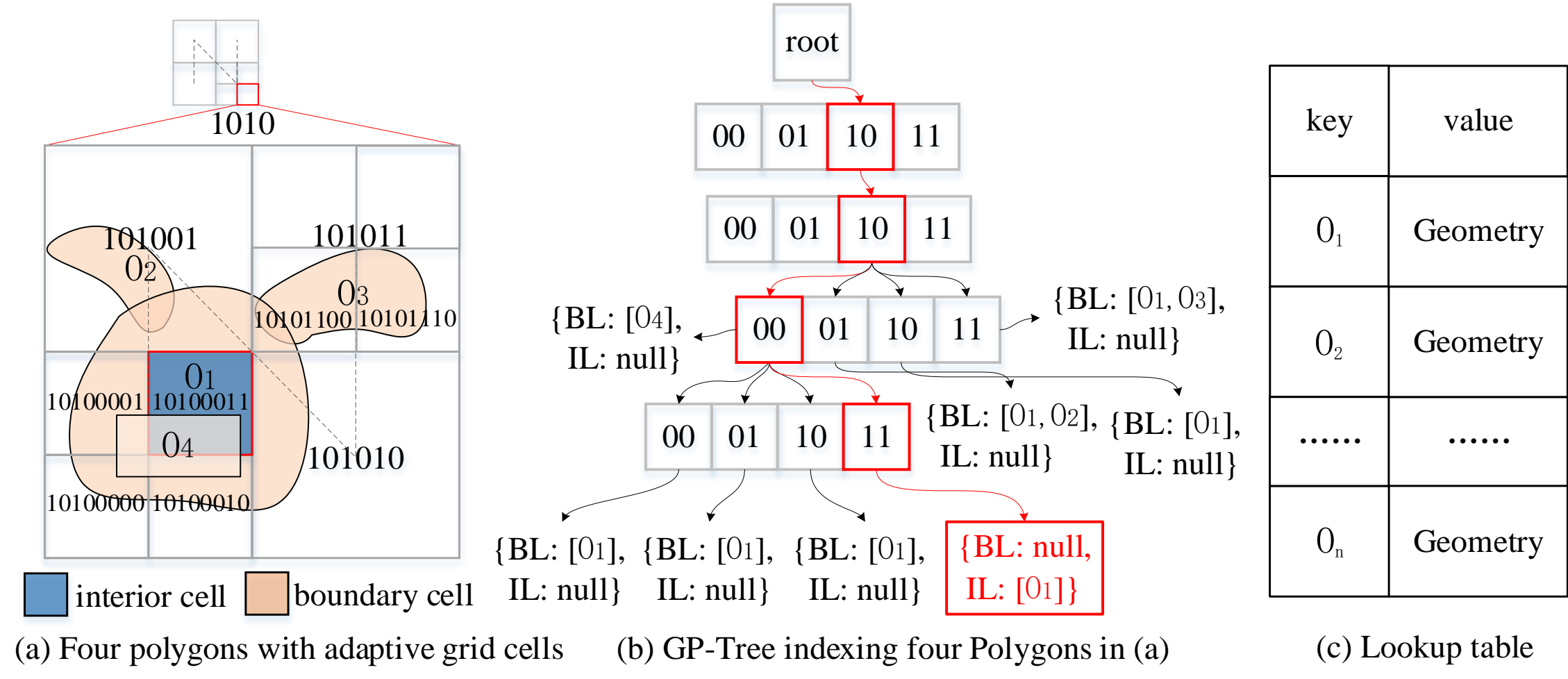

(a) Four polygons with adaptive grid cells (b) GP-Tree indexing four Polygons in (a) (c) Lookup table

Figure 3. The construction of GP-Tree.

Due to the hierarchical organization of adaptive grid cells, objects may reside at arbitrary tree levels. Therefore, both intermediate nodes and leaf nodes maintain object IDs associated with the corresponding cell regions. Specifically, each node stores a node value defined as:

$$node_{value} = \{BL{:}\, Long[\quad],IL{:}\, Long[\quad]\} \tag{1}$$

Where $BL$ and $IL$ denote the boundary list and interior list, respectively. Both lists store only the IDs (Long type) of spatial objects associated with boundary cells and interior cells. For example, since “10100011” is an interior cell of $O_1$ in Figure 3(a), the ID of $O_1$ is recorded in the $IL$ of the corresponding node.

In addition to the prefix tree, GP-Tree maintains a lookup table that maps object IDs to their corresponding geometries, as illustrated in Figure 3(c). The lookup table serves as the object-reference structure, enabling efficient retrieval of complete geometries during query refinement. Algorithm 1 presents the pseudocode of basic GP-Tree construction.

**Algorithm 1: Construction of basic GP-Tree**

```
Input: S // spatial dataset with approximated grid cells
Output: GPTree
1.  GPTree←new GPTree();
2.  lookupTable←new HashMap();
3.  for s in S do
4.     lookuptable.put(s.id, s); // store spatial objects
5.     for cell in s.cells do
6.        currentNode←GPTree.rootNode;
7.        currentLevel = 0;
8.        while (currentLevel < cell.level) do
9.           bits←getBits(cell.encoding, currentLevel); // retrieve 2-bit encoding at the current level
10.          currentNode←currentNode.subNode[bits]; // access the child node indexed by bits (0–3)
11.          currentLevel++;
12.       end while
13.       if (cell.isInterior) do currentNode.IL.add(s.id); // store object IDs for interior cells
14.       else do currentNode.BL.add(s.id); // store object IDs for boundary cells
15.       end if
16.    end for
17. end for
18. return GPTree;
```

### *4.2. Optimization of GP-Tree*

GP-Tree requires further optimization for two primary reasons: (1) Due to the hierarchical organization of grid cells, object references in the GP-Tree may reside at arbitrary tree levels. Consequently, intermediate nodes are required to maintain object-reference lists, increasing memory consumption, while query processing must inspect references at intermediate tree levels during traversal, introducing additional query overhead. (2) These grid cells cluster under specific nodes of the prefix tree, rendering the constructed index tree sparse. As shown in Figure 3(b), both the first and second levels of the tree have only one valid child node (forming a single path), with these nodes storing no values. Such query-ineffective layers elongate the tree height and thus the query path. To address these issues, this section introduces data node optimization and a pruning strategy to enhance the overall efficiency of GP-Tree.

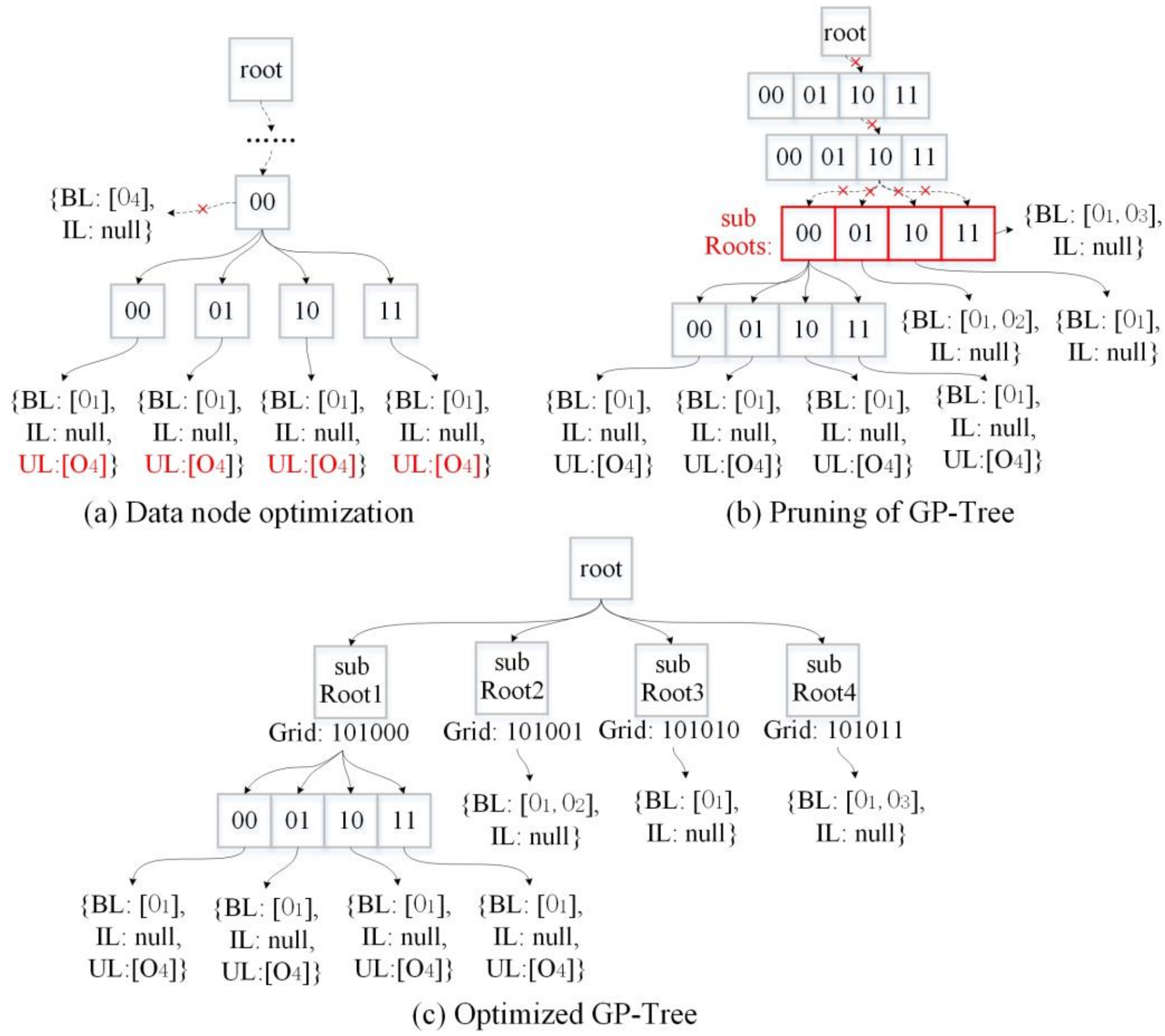

Figure 4. The optimization processes of GP-Tree.

**Node optimization.** The objective of node optimization is to eliminate the memory and query overhead caused by storing object references at arbitrary tree levels. To achieve this, object references maintained in intermediate nodes are migrated to their corresponding leaf nodes, ensuring that all spatial object references are stored exclusively in leaf nodes.

Specifically, the references in the IL of a non-leaf node are propagated to the ILs of all its descendant leaf nodes, as illustrated in Figure 4(a). This operation is safe because if a grid cell is an interior cell of a spatial object, all of its child cells are likewise interior cells of that object. However, when a non-leaf node holds references in its BL, direct propagation becomes ambiguous, as the spatial object may intersect only a subset of the child cells (nodes). To address this, GP-Tree introduces an uncertain list (UL) in leaf nodes, which stores references inherited from the parent's BL. During query processing, if a UL is accessed, refinement is required to verify the actual spatial relationship. This approach ensures data accuracy while reducing the number of nodes that store references, thereby optimizing in-memory usage. Figure 4(a) shows this process, and Algorithm 2 provides the corresponding pseudocode.

**Algorithm 2: Function nodeOptimization(root)**

```
1.  Queue<Node> queue←new PriorityQueue<>();
2.  queue.offer(root);
3.  while (!queue.isEmpty()) do
4.     Node node←queue.poll();
5.     if (node is null or node.isLeafNode) do
6.        continue;
7.     else if (node.hasItem) do
8.        for (i:=0 to 3) do
9.           subNode←node.subNode[i];
10.          subNode.IL.merge (node.IL);
11.          subNode.UL.merge(node.BL);
12.          subNode.UL.merge(node.UL);
13.          node.clear();
14.          queue.offer(subNode);
15.       end for
16.    end if
17. end
18. return root
```

**Pruning Strategy.** Due to the sparsity of GP-Tree, many upper-level nodes do not store any spatial references, resulting in redundant tree height and longer query paths. To address this issue, GP-Tree introduces a pruning strategy that reduces tree height by merging sparse subtrees, as illustrated in Figure 4(b). The pseudocode of this strategy is provided in Algorithm 3. The core idea is to treat each direct child of the root node as a sub-root and iteratively determine whether it can be merged with its valid child nodes. Specifically, all sub-root nodes are added to a processing queue. For each sub-root, its valid children are examined to assess whether they can be merged upward to form new sub-root nodes. If the total number of valid sub-roots under the root remains fewer than four after merging, the merge is considered successful, and newly formed sub-roots are re-added to the queue for further pruning. Otherwise, the current sub-root remains unchanged and is removed from the queue. This process is repeated recursively until no further sub-root nodes can be merged. Figure 4(c) shows the optimized GP-Tree structure. This strategy eliminates redundant layers, reduces the height of the tree, and ultimately enhances the retrieval performance of GP-Tree.

**Algorithm 3: Function pruning(root)**

```
1.  Queue<subRoot> queue←new PriorityQueue<>();
2.  subNodes←root.subNodes;
3.  for node in subNodes
4.     if (node!=null) do
5.        subRoot←createSubRoot(node);
6.        root.subRoot.add(subRoot);
7.        queue.offer(subRoot);
8.     end if
9.  end for
10. unMergeRootSize = 0;
11. subRootSize = queue.size+unMergeRootSize;
12. while (subRootSize ≤ 4 and queue !=null) do
13.    subRoot←queue.poll();
14.    if (!subRoot.isLeafNode) do
15.       preMergedSize = subRoot.validSubNodes.size;
16.       if (subRootSize+preMergedSize-1) > 4 do
17.          unMergeRootSize++;
18.       else do
19.          subRoots←mergeSubRoots(subRoot, node.validSubNodes);
20.          queue.offerAll(subRoots);
21.       end if
22.       subRootSize = queue.size+unMergeRootSize;
23.    end if
24. end while
25. return root
```

### *4.3. Complexity analysis of GP-Tree construction*

The construction process of the GP-Tree employs fine-grained grid cells instead of coarse MBRs, while the search process uses bitwise operations on cell encoding rather than geometric computations on MBRs. The runtime complexity of a prefix search is $O(k)$, where $k$ is the length of cell encoding, compared to the $O(log(n))$ for MBR-based spatial indexes, where $n$ is the number of spatial objects. In other words, the number of node accesses in GP-Tree is bounded by the maximum key length $k_{max}$, which is 60 when 30 quadtree levels are used. In practice, a smaller $k_{max}$ is often sufficient for spatial objects. Therefore, the GP-Tree significantly improves retrieval efficiency compared with traditional MBR-based tree structures.

## 5. Spatial queries based on GP-Tree

In this section, a variety of spatial query operations are implemented on top of GP-Tree, including range query, $\varepsilon$-distance query, and $k$NN query.

### *5.1. Range query*

A range query is performed to retrieve all spatial objects located within a given region. The formal definition of the range query is as follows.

**Definition 1. (Range Query).** Given a query range $q$, a dataset $S$, and operator $\Theta$ (intersect or contain), a range query returns all spatial objects $s$ in $S$ that intersect or lie in the query range $q$, i.e., $\Theta(s,q)$ is true. Formally,

$$RangeQuery(q,S,\Theta) = \{ s \mid s \in S \wedge \Theta(s,q) \} \tag{2}$$

The query processing of the GP-Tree is divided into three steps as illustrated in Figure 5. Algorithm 4 presents the pseudocode of the range query processes.

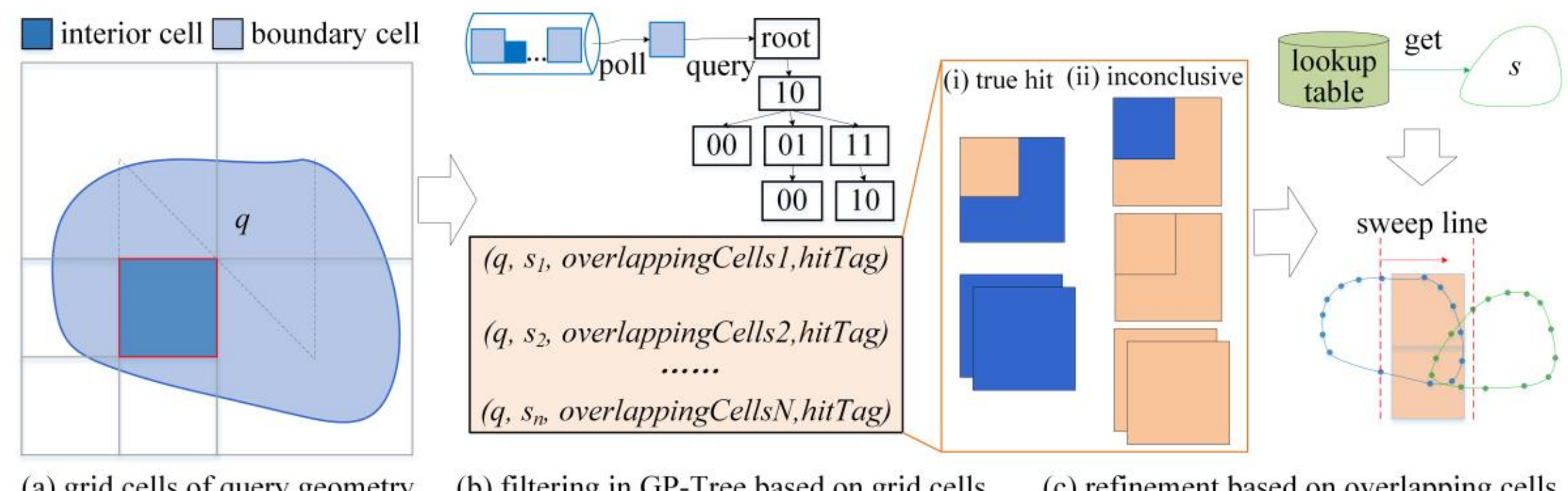

Figure 5. The pipeline of range query in GP-Tree.

**Step 1: Rasterization of the query object** (line 1). As shown in Figure 5(a), the query object $q$ is first rasterized into a set of multi-type grid cells following the method in Section 3.3.

**Step 2: Filtering based on GP-Tree.** The filtering step finds all spatial objects $s$ that have overlapping cells with $q$ and records their overlapping cells. Specifically, for each cell of $q$, its encoding is used to perform a prefix search in the GP-Tree. This search process proceeds in a top-down manner, consuming two bits of the cell encoding at each level, and is similar to the GP-Tree construction procedure (lines 3-13). If the query cell encoding has been fully consumed and the current node still has children, all descendant nodes of the current node and their references are traversed and collected as candidate matches (line 15). Each candidate match is stored as a tuple $(q, s_{id},\ overlappingCells,$ $hitTag)$. As shown in Figure 5(b), the filtering result falls into two cases: (i) true hit, where the candidate match definitely intersects if any of the overlapping cells is an interior cell; (ii) uncertain, where the intersection cannot be determined solely based on the overlapping cells. In case (i), the time-consuming refinement can be omitted.

**Algorithm 4: range query processes of GP-Tree**

```
Input: GPTree gpTree, lookupTable, query range q
Output: result // List<Geometry>
1.  GridGeometry gridQ←new GridGeometry(q);
2.  root←gpTree.root;
    // filtering
3.  for cell in gridQ.cells do
4.      queryLevel←cell.level;
5.      for subRoot in root.subRoots do
6.          if subRoot.contained(cell) do
7.              currentNode←subRoot;
8.              currentLevel←currentNode.level;
9.              break;
10.     while (currentNode is not leaf node and currentLevel ≤ queryLevel) do
11.         pointer←getPointer(cell, currentLevel);
12.         currentNode←currentNode.subNode[pointer];
13.         currentLevel++;
14.     if currentNode is leaf node: visitorItem(currentNode, itemVisitor);
15.     else: broadFirstSearch(currentNode, itemVisitor);
    // refinement
16. for (q,s_id,overlappingCells,hitTag) in itemVisitor do
17.     s←lookupTable.get(s_id)
18.     if (hitTag is true) then
19.         result.add(s)
20.     else then
21.         s_part←s.intersect(overlappingCells);
22.         q_part←q.intersect(overlappingCells);
23.         if sweepLine(q_part, s_part) then
24.             result.add(s);
25.         end if
26.     end if
27. end for
28. return result;
```

**Step 3: Refinement based on overlapping grid cells.** For uncertain candidate matches, GP-Tree performs exact geometric refinement restricted to the overlapping grid cells to confirm spatial relationships. Because the spatial relationship between a query object and a candidate depends only on their overlapping cells, the refinement can ignore most of each object's geometry segments and operate solely on the segments that intersect those cells, as shown in Figure 5(c), which greatly reducing computational overhead. Specifically, the corresponding geometries are first retrieved from the lookup table using their IDs. For each geometry, only the segments intersecting with the overlapping cells are extracted. The spatial relationship of a candidate match is then determined by applying the sweep line algorithm to these segments (lines 21–24).

**Complexity analysis of range query in GP-Tree.** The query process involves two major steps: tree traversal for candidate retrieval and geometric refinement for precise verification. As analyzed in Section 4.3, the retrieval efficiency of GP-Tree is significantly higher than that of MBR-based spatial indexes. The refinement in traditional indexes performs exact geometric computations on all vertices of both the retrieved candidate and the query object, resulting in a time complexity of $O(N\,log(N))$ where $N$ is the total number of vertices. In contrast, GP-Tree restricts refinement to the $k$ overlapping grid cells. Letting $n$ denote the average number of segments per cell ($n \ll N$), the refinement complexity in GP-Tree is $O(k \cdot n\,log(n))$, which is typically much smaller than $O(N\,log(N))$. Overall, from a theoretical perspective, GP-Tree achieves superior range query performance compared with traditional spatial indexes.

### *5.2. ε-Distance query*

The ε-distance query retrieves all spatial objects whose distances to the query object are within a given threshold ε. Its formal definition is as follows.

**Definition 2: ($\varepsilon$-distance Query).** Given a query spatial object $q$, a dataset $S$, and a distance threshold $\varepsilon$, a distance query returns all spatial objects $s$ in $S$ whose distance to $q$ is at most $\varepsilon$. Formally,

$$\varepsilon DistanceQuery(q, S, \Theta) = \{ s \mid s \in S \wedge \|q, s\| \leq \varepsilon \} \tag{3}$$

The $\varepsilon$-distance query in GP-Tree is designed to efficiently identify spatial objects whose distance to a query object is within a given threshold $\varepsilon$. The method follows several key steps to leverage the fine-grained grid-based structure of GP-Tree, as illustrated in Figure 6. Algorithm 5 presents the pseudocode of the $\varepsilon$-distance query processes.

**Step 1: Extending grid cells of the query object** (lines 1-3). Similar to the range query process, the query object $q$ is first represented as a set of grid cells according to the method described in Section 3.3. Each cell of $q$ is then expanded by a distance of $\varepsilon$ to generate a set of expanded cells, as shown in Figure 6(a). The newly generated cells are considered boundary cells. If the diagonal length of a boundary grid cell of $q$ is smaller than $\varepsilon$, the cell is converted to an interior cell after expansion, as shown in Figure 6(b).

**Step 2: Merging cells** (line 4). In order to reduce the total number of cells, adjacent cells of the same type are merged into their parent cell, as shown in Figure 6(c). To reduce the number of grid cells, adjacent cells of the same type that share the same parent are merged into their parent cell, as illustrated in Figure 6(c). Specifically, when four cells that belong to the same parent grid are adjacent and of identical type (either interior or boundary), they are replaced by their parent grid cell. This hierarchical merging process simplifies the grid representation while maintaining the spatial integrity of the query object.

**Step 3: Filtering and refinement of distance query** (lines 5-15). With the help of grid-based approximations, the $\varepsilon$-distance query is transformed into a range query, as illustrated in Figure 6(d). Specifically, each grid cell performs a range query using the GP-Tree. Similar to Section 5.1, since grid cells are divided into interior and boundary cells, the filtering results of distance queries also fall into two cases: true hit and uncertain. Candidate objects retrieved from interior cells of $q$ are all true hit objects, thus eliminating the need for the time-consuming distance refinement process. For each uncertain candidate, its actual distance to $q$ is computed. If the distance is less than $\varepsilon$, the candidate is considered a valid result, otherwise, it is discarded.

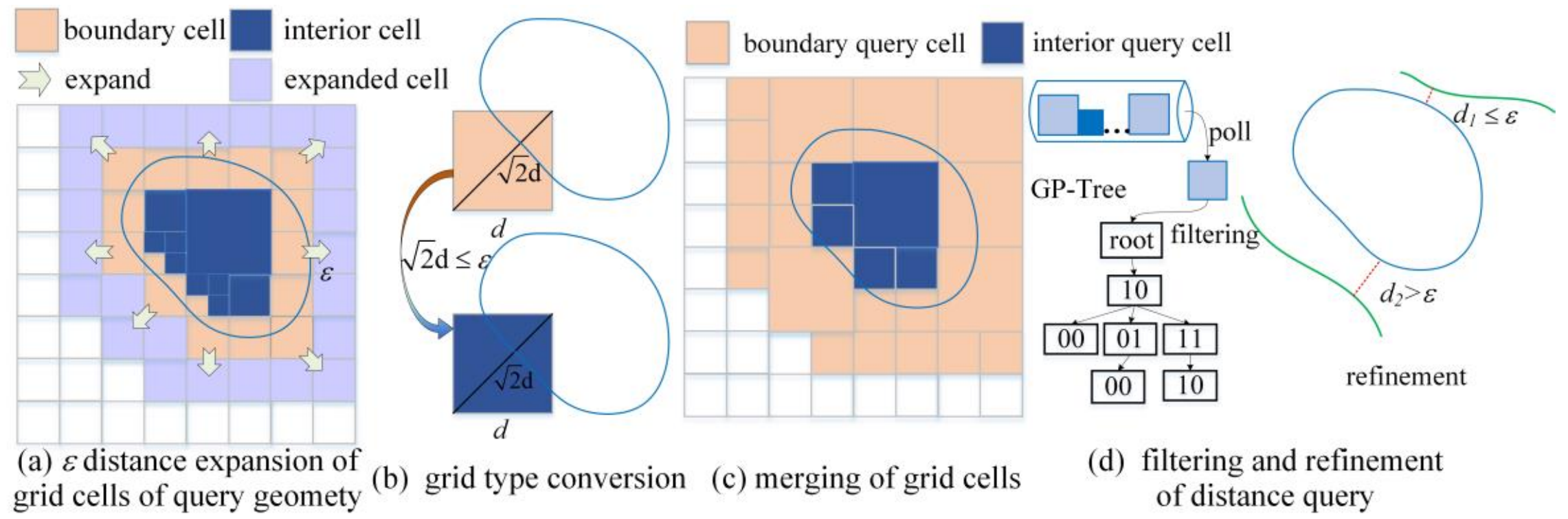

Figure 6. The pipeline of $\varepsilon$-distance query in GP-Tree.

By performing a distance expansion on the query object's approximate grid cells, the $\varepsilon$-distance query is converted into an equivalent range query, enabling inexpensive prefix searches on cell encodings instead of direct geometric distance computations. As shown in Section 4.3, GP-Tree's prefix searches rapidly retrieve candidates, and only those intersecting the expanded cells proceed to further processing. Moreover, the filtering stage produces a portion of true-hit candidates, which further reduces the number of objects requiring exact distance computation and thus improves average query throughput.

**Algorithm 5: $\varepsilon$–distance query processes of GP-Tree**

**Input:** *GPTree gpTree, lookupTable, query object q, distance* $\varepsilon$
**Output:** *result // List<Geometry>*
1. *gridQ←gridGeometry(q);*
2. *extendCells←cellExtend(gridQ.getCells,* $\varepsilon$*);*
3. *convCells←cellConv(q.getCells,* $\varepsilon$*);*
4. *queryCells←MergeCells(extendCells, convCells);*
5. **for** *cell* **in** *queryCells* **do**
6.     *candidateSet←gpTree.rangequery(cell);*
7.     **for** *s* **in** *candidateSet* **do**

```
8.          if cell.tag is interior do
9.             result.add(s); // true hit
10.         else if q.distance(s) ≤ ε do
11.            result.add(s);
12.         else {continue;} // false hit
13.         end if
14.      end for
15.   end for
16.   return result;
```

### 5.3. KNN query

A kNN query identifies the $k$ nearest spatial objects to a given query object. The formal definition of the kNN query is as follows.

**Definition 3: (kNN Query).** Given a query spatial object $q$, a spatial dataset $S$, and a positive integer $k$, a kNN query returns a set of spatial objects $S' \subseteq S$, where $|S'| = k$, and for each $s \subseteq S'$, $o \in S \setminus S'$, $\|q, s\| \leq \|q, o\|$. Formally,

$$kNNQuery(q, S, k) = \left\{ s \in S' \,\middle|\, \begin{array}{c} S' \subseteq S \wedge |S'| = k \\ \wedge\, \forall o \in S \setminus S', \|q, s\| \leq \|q, o\| \end{array} \right\} \quad (4)$$

To improve the efficiency of kNN queries, a grid histogram-based secondary index (GHSI) is proposed. As illustrated in Figure 7(a), GHSI captures the distribution of spatial objects by dividing the space into fixed-level grid cells, with each cell recording the number of spatial objects it contains. For non-point objects that may intersect multiple cells of GHSI, only the cell containing the object's center point is recorded. This histogram information is stored in a HashMap, where the key represents the cell ID, and the value indicates the number of objects within that cell.

Leveraging GHSI, the kNN query process in GP-Tree iteratively expands the query cells containing the query object in an inside-out manner until the k nearest neighbors are found. This process comprises three steps, as shown in Figure 7: (1) query cell extension, (2) rough kNN query, (3) refinement. Algorithm 6 provides the pseudocode for the kNN query process.

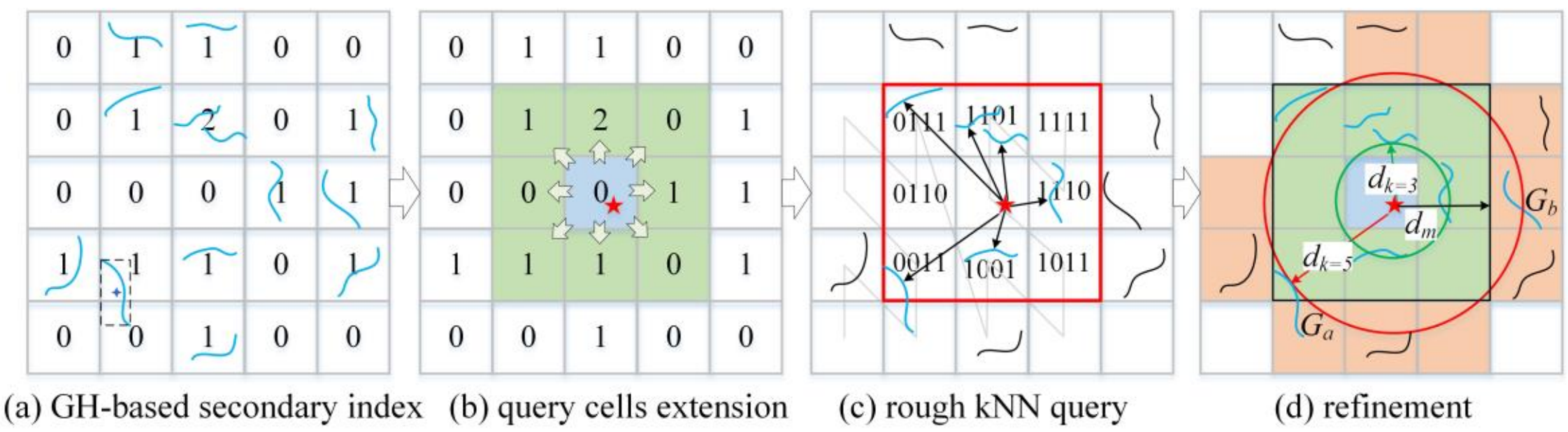

Figure 7. The pipeline of kNN query in GP-Tree.

**Step 1: Query cell extension** (lines 1–7). As illustrated in Figure 7(b), the query object $q$ (marked as a red pentagram) is first approximated by grid cells in GHSI to form the initial query cells. If the number of objects within these cells is less than $k$, adjacent cells surrounding the current query region are added as new query cells. This expansion continues iteratively until the total number of objects within the query cells exceed $k$.

**Step 2: Rough kNN query** (lines 8–12). The query cells obtained in Step 1 are used to perform range queries on the GP-Tree. The distance to $q$ is calculated for each candidate object retrieved by range queries, and all candidates are sorted in ascending order of distance. The top $k$ objects are selected as the rough kNN query results.

Since the query boundary formed by query cells is rectangular rather than circular, the rough kNN query results may lack absolute accuracy. To validate the results, the distance $d_k$ (the distance from $q$ to the $k$-th nearest object in rough kNN query results) is compared with $d_m$, which represents the minimum distance from $q$ to query boundary. If $d_k \leq d_m$, the rough kNN results are guaranteed to be accurate, and no further refinement is necessary. This is illustrated in Figure 7(d) for the case of $k = 3$, where $d_k = 3$ (the green line) is less than $d_m$. Conversely, when $d_k > d_m$, refinement is necessary, as demonstrated in Figure 7(d) for $k = 5$, where $d_k = 5$ (the red line) exceeds $d_m$.

**Step 3: Refinement** (lines 13–17). During refinement, a circle centered at the centroid of $q$ with a radius of $d_k$ is generated, as shown in Figure 7(d). Range queries are then performed using the orange grid cells that intersect the circle but have not yet been retrieved. If the newly retrieved object is found to be closer to $q$ than $d_k$, both the kNN result and $d_k$ are updated accordingly. The refinement process continues until all orange cells have been queried.

**Algorithm 6**: **kNN query process**

```
Input: GHSI, GPTree, a query object q, the number k
Output: TreeSet<Geometry> kNNResult
1.   queryCells ← GHSI.getCells(q);
2.   kNNResult ← new TreeSet<>(Geometry);
3.   count ← 0;
4.   while count < k:
5.      count ← GHSI.count(queryCells);
6.      queryCells ← GHSI.gridExtension(queryCells);
7.   end while
8.   queryResults ← GPTree.rangeQuery(queryCells);
9.   for result in queryResults:
10.     kNNResult.add(result);
11.  end for
12.  d_k ← kNNResult.get(k-1).getDistanceToPoint();
13.  unViewedCells ← GHSI.generateUnviewedCells(q, d_k, queryCells);
14.  unViewedResults ← GPTree.rangeQuery(unViewedCells);
15.  for result in unViewedResults:
16.     kNNResult.add(result)
17.  end for
18.  return kNNResult.take(k)
```

**Complexity analysis of kNN query in GP-Tree.** In GP-Tree, kNN queries are transformed into range queries by employing the auxiliary GHSI structure. GHSI is built by a single pass over the dataset, so its construction cost is $O(N)$, where $N$ is the total number of spatial objects. The GHSI is implemented using a HashMap, allowing both insertion and lookup operations to be performed in expected $O(1)$ time. If the GHSI grid level is $l$, the space is divided into $4^l$ grid cells and the storage space occupied by GHSI is $4^l \times 12$ bytes (each entry uses 8 bytes for the cell id and 4 bytes for the count). For example, when $l = 11$, the storage overhead is only on the order of tens of megabytes. Furthermore, by transforming kNN queries into equivalent range queries, GP-Tree leverages bitwise prefix searches on grid encodings, replacing computationally expensive geometric distance computations.

## 6. Experiments and analysis

### *6.1. Datasets and settings*

**Datasets.** To evaluate the performance of GP-Tree, several datasets from UCR STAR (https://star.cs.ucr.edu/) are employed in the below experiments: (1) Tweets, a point dataset, consists of approximately 20 million geotagged tweets; (2) POIs, another point dataset with about 25 million points of interest across the United States; (3) WaterL, a linestring dataset consisting of 5.5 million linear water features; (4) Roads is also a linestring dataset comprising around 18 million road segments throughout the U.S.; (5) WaterP, a polygon dataset composed of approximately 2.3 million boundaries of water area; (6) Buildings is a polygon dataset with more than 20 million building footprints in the United States; (7) Administrative Boundaries is a polygon dataset representing multi-level administrative regions, including COUNTY (city level), COUSUB (county subdivision), and BG (block group), each containing 1,000 sampled boundary objects used as query ranges; and (8) RandomP is a dataset of 1000 randomly generated point objects, used for k-NN queries in the experiments. Detailed information about these datasets is summarized in Table 3.

Table 3. Information of experimental datasets.

| **datasets** | **dataset purpose** | **geometry type** | **data size** | **number of records** | **total number of points** | **average number of points** |
|---|---|---|---|---|---|---|
| Tweets | indexed dataset | Point | 1.5 GB | 19.8 M | 19.8 M | 1 |
| POIs | | Point | 1.0 GB | 25.1 M | 25.1 M | 1 |
| Roads | | LineString | 11.0 GB | 18.0 M | 345.4 M | 19 |
| WaterL | | LineString | 8.6 GB | 5.5 M | 284.3M | 51 |
| Buildings | | Polygon | 9.5 GB | 22.0 M | 176.2 M | 8 |
| WaterP | | Polygon | 2.6 GB | 2.3M | 97.7 M | 43 |
| COUNTY | query scope | Polygon | \ | 1000 | 2.5 M | 2494 |
| COUSUB | | Polygon | \ | 1000 | 0.8 M | 817 |
| BG | | Polygon | \ | 1000 | 0.3 M | 305 |
| RandomP | | Point | \ | 1000 | 1000 | 1 |

**Settings.** Table 4 summarizes the related parameters in the experiments, with the default values shown in bold. All experiments are conducted on a single-node machine running Ubuntu 18.04, equipped with a 12-core CPU, 64 GB of RAM, and a 512 GB disk.

Table 4. Parameter settings.

| **Parameters** | **Settings** |
|---|---|
| indexed dataset | Tweets, POIs, WaterL, **Roads**, WaterP, Buildings |
| query scope | COUNTY, COUSUB, **BG**, RandomP |
| percentage of datasets (%) | 10, 20, 30, 40, 50, 60, 70, 80, 90, **100** |
| $\varepsilon$ in distance query | 0.01, 0.02, **0.03**, 0.04, 0.05, 0.06, 0.07, 0.08, 0.09, 0.10 |
| $k$ in kNN query | 1, 5, 10, **20**, 30, 40, 50 |
| *SEG* | 10, 15, **20**, 25, 30, 35 |

**Baseline methods.** The performance of GP-Tree is compared with four baseline methods: STR-Tree(Leutenegger *et al.* 1997), B+Tree(Comer 1979), MultiR-Tree(Schoemans *et al.* 2024) and Quad-Tree. STR-Tree and Quad-Tree are traditional singe-entry spatial indexes built on MBRs, which are widely used in spatial data systems such as SpatialHadoop(Eldawy and Mokbel 2015) and GeoSpark(Yu *et al.* 2015). B+Tree is a self-balancing binary search tree commonly used in NoSQL databases (such as HBase). In this paper, we use the encoding of the grid cells of spatial objects as keys and organize them into an in-memory-based B+Tree (https://github.com/Morgan279/MemoryBasedBPlusTree). MultiR-Tree is a multi-entry version of R-Tree, where each non-point spatial object is represented as multiple MBRs and these MBRs are organized into an R-Tree.

**Metrics.** We focus on three metrics: (1) **Throughput**, defined as the number of query requests completed per minute, serves as a metric for evaluating the query efficiency of various spatial indexes; (2) **Hit rate**, which is a metric for evaluating filtering capability of spatial indexes, including true hit rate (THR) defined as $n_t/n_c$, false hit rate (FHR) defined as $n_f/n_c$, uncertain rate (UCR) defined as $n_u/n_c$, where $n_t$ is the number of true hits, $n_f$ is the number of false hits and $n_c$ is the number of the candidate pairs whose MBRs intersect; (3) **In-memory cost** is used to evaluate spatial indexes size.

## *6.2. GP-Tree vs. baselines*

### *6.2.1. Range query*

As illustrated in Figure 8, GP-Tree is compared with STR-Tree, B+Tree, MultiR-Tree and Quad-Tree on the range query performance.

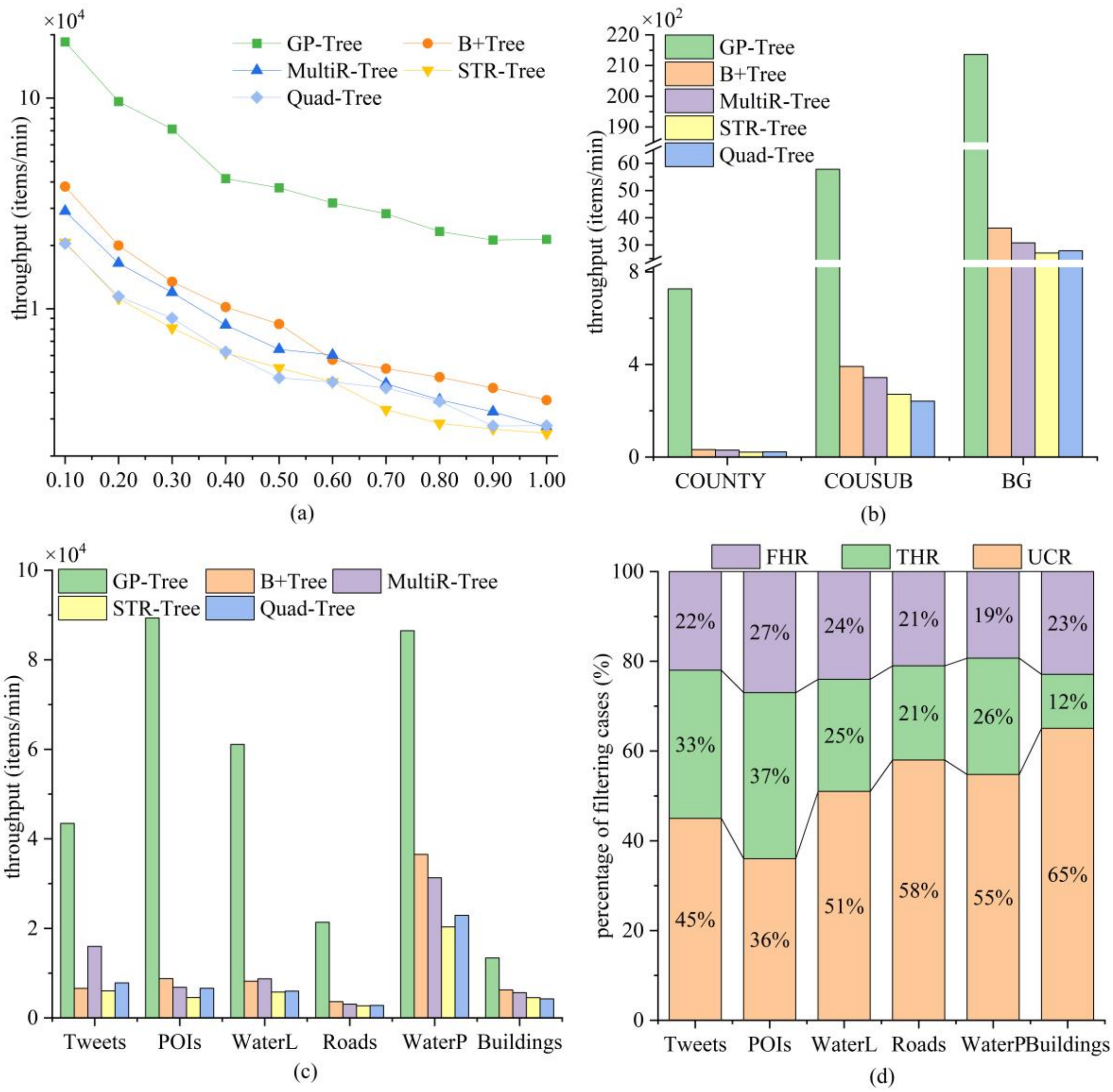

Figure 8. Comparison of different spatial indexes in range queries on (a) different dataset sizes, (b) different query scopes and (c) different datasets. (d) The filtering ability of GP-Tree across datasets.

**Impact of dataset sizes.** To evaluate scalability, the size of the Roads dataset is gradually increased from 10% to 100%. We then build different spatial indexes on these datasets and conduct range query experiments using the BG administrative boundaries. As shown in Figure 8(a), the throughput of all indexes decreases as the dataset size grows. This is because larger dataset leads to deeper and heavier index structures and introduces more candidate objects, resulting in increased filtering and refinement costs. Across all scales, GP-Tree consistently maintains a substantial performance advantage, achieving speedups of approximately 6.13× over MultiR-Tree, 5.03× over B+Tree, 8.02× over STR-Tree, and 7.55× over Quad-Tree.

**Impact of query ranges.** To assess the effect of query-range size, we build spatial indexes on full Roads dataset and execute range queries using administrative boundaries at different levels. As shown in Figure 8(b), the throughput of all indexes increases as the query range shrinks from COUNTY level to BG level, because smaller query ranges retrieve fewer candidate objects. Across all query ranges, GP-Tree consistently delivers the best performance, achieving speedups of over 6.95×–24.20× MultiR-Tree, 5.90×–22.69× over B+Tree, 7.91×–34.57× over STR-Tree, and 7.67×–23.98× over Quad-Tree.

**Impact of dataset types.** Range-query experiments are then conducted on various spatial datasets using the BG dataset as query objects. GP-Tree consistently achieves the highest throughput across all datasets, as shown in Figure 8(c). Specifically, on point datasets (Tweets and POIs), GP-Tree outperforms MultiR-Tree, B+Tree, STR-Tree, and Quad-Tree by factors of approximately 7.91×, 8.40×, 13.45×, and 9.54×, respectively. On linestring datasets (WaterL and Roads), the performance gains reach 6.95×, 6.70×, 9.25×, and 8.96×. On polygon datasets (WaterP and Buildings), GP-Tree improves performance by 2.60×, 2.25×, 3.65×, and 3.46×. These results indicate that GP-Tree achieves larger performance gains on point and linestring datasets than on polygon datasets, primarily because its filtering effectiveness declines for polygonal objects. To quantify this effect, we further analyze GP-Tree's filtering capability across datasets (Figure 8(d)). The UCR on point and linestring datasets is 36%–58%, substantially lower than the 55%–65% observed on polygon datasets, which explains the difference in performance improvements.

*6.2.2. $\varepsilon$-Distance query*

Figure 9 compares the $\varepsilon$-distance query performance of GP-Tree against STR-Tree, B+Tree, MultiR-Tree and Quad-Tree.

**Different data sizes**. Similar to the range-query experiments, we build spatial indexes on the Roads dataset at varying scales (10%–100%) and execute $\varepsilon$-distance queries ($\varepsilon$=0.03°) using BG boundaries. As shown in Figure 9(a), throughput of all indexes decreases as dataset size grows, which is likewise caused by larger index structures and a growing number of candidate objects as in the range-query experiments. GP-Tree consistently achieves the best performance across all scales, outperforming MultiR-Tree, B+Tree, STR-Tree, and Quad-Tree by approximately 6.07×, 3.04×, 7.41×, and 6.71×, respectively.

**Different $\varepsilon$ values.** Figure 9(b) illustrates the throughput of $\varepsilon$-distance queries on the Roads dataset as $\varepsilon$ varies from 0.01 to 0.10. Obviously, throughput decreases with increasing $\varepsilon$ for all indexes, because a larger $\varepsilon$ expands the query scope, producing more candidates. GP-Tree consistently outperforms the baselines, achieving average speedups of 6.27×, 3.47×, 7.65×, and 7.01 over MultiR-Tree, B+Tree, STR-Tree, and QuadTree, respectively.

**Different dataset types.** Distance queries with $\varepsilon$=0.03 are evaluated across various spatial datasets using BG boundaries as query objects. As shown in Figure 9(c), GP-Tree attains the highest throughput on all datasets. The speedups over baseline spatial indexs are 3.06×–4.52× for point datasets (Tweets, POIs), 3.24×–11.79× for linestring datasets (WaterL, Roads), and 5.36×–25.27× on polygon datasets (WaterP, Buildings). The performance improvement of GP-Tree in $\varepsilon$-distance queries is partly attributable to the filtering capability of adaptive grid cell approximations, which effectively reduce the number of candidate objects forwarded to the refinement phase, as shown in Figure 9(d).

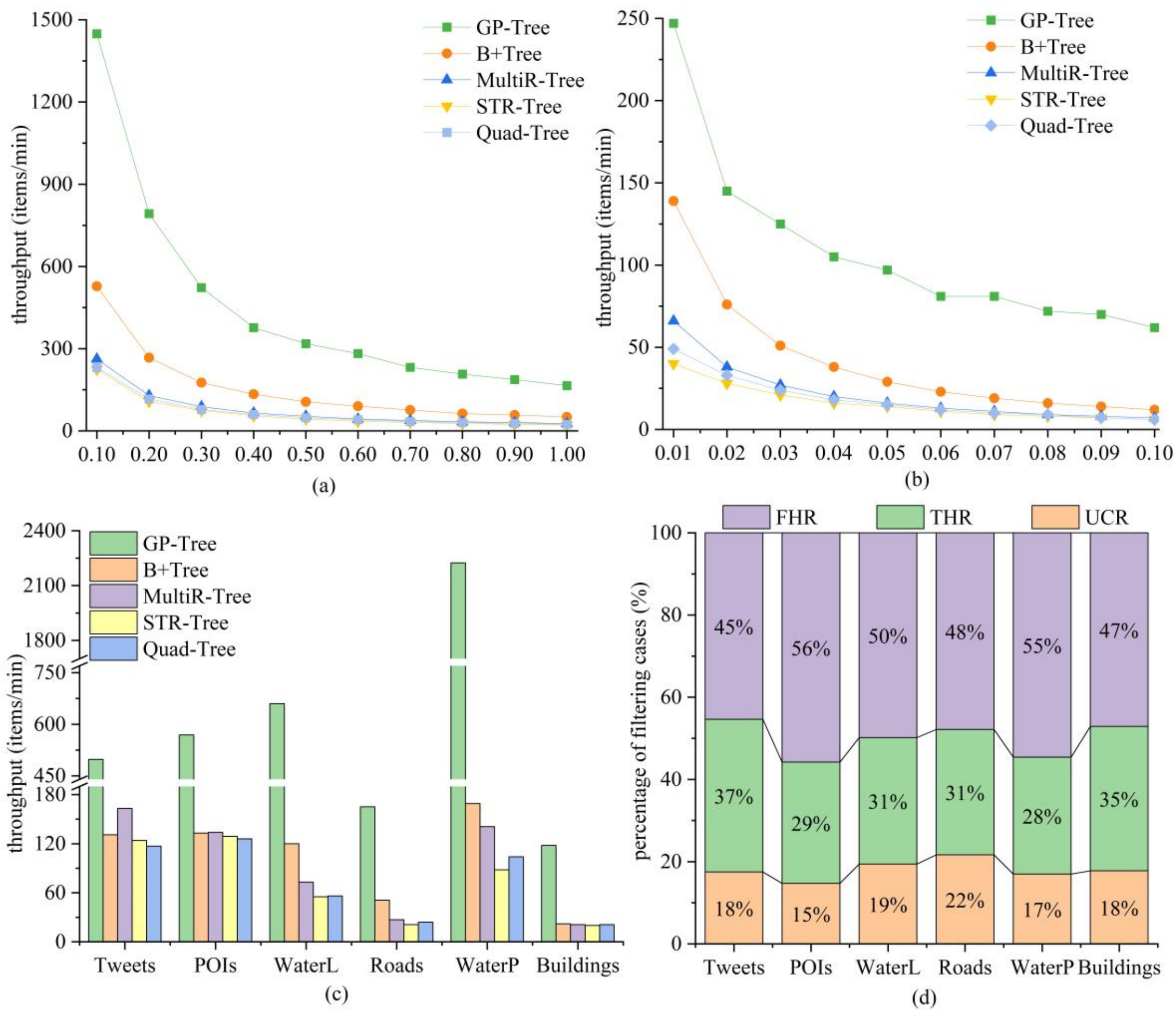

Figure 9. Comparison of different spatial indexes in $\varepsilon$-distance queries on (a) different dataset sizes, (b) different $\varepsilon$ values and (c) different datasets. (d) The filtering ability of GP-Tree across datasets.

### *6.2.3. kNN query*

Figure 10 compares the kNN query performance of GP-Tree with STR-Tree, B+Tree, MultiR-Tree. Quad-Tree is excluded from this evaluation as it does not support kNN queries.

**Impact of dataset sizes**. We vary the size of the Roads dataset from 10% to 100% and perform kNN queries using the RandomP dataset with $k$=20. Figure 10(a) shows that the throughput of all indexes decreases with increasing datasets, as larger datasets lead to deeper index trees and accordingly higher traversal overhead. GP-Tree outperforms all baseline indexes when the dataset size exceeds 80% and exhibits the smallest performance degradation overall. Specifically, as the dataset size increases from 10% to 100%, the throughput of GP-Tree decreases by only 43.6%, compared with 58.0% for B+Tree, 74.1% for MultiR-Tree, and 70.9% for STR-Tree. This indicates that GP-Tree is more suitable for large-scale datasets. The advantage stems from its prefix-tree–based organization with $O(l)$ search complexity, where $l$ denotes the max encoding length, making query performance less sensitive to data volume growth.

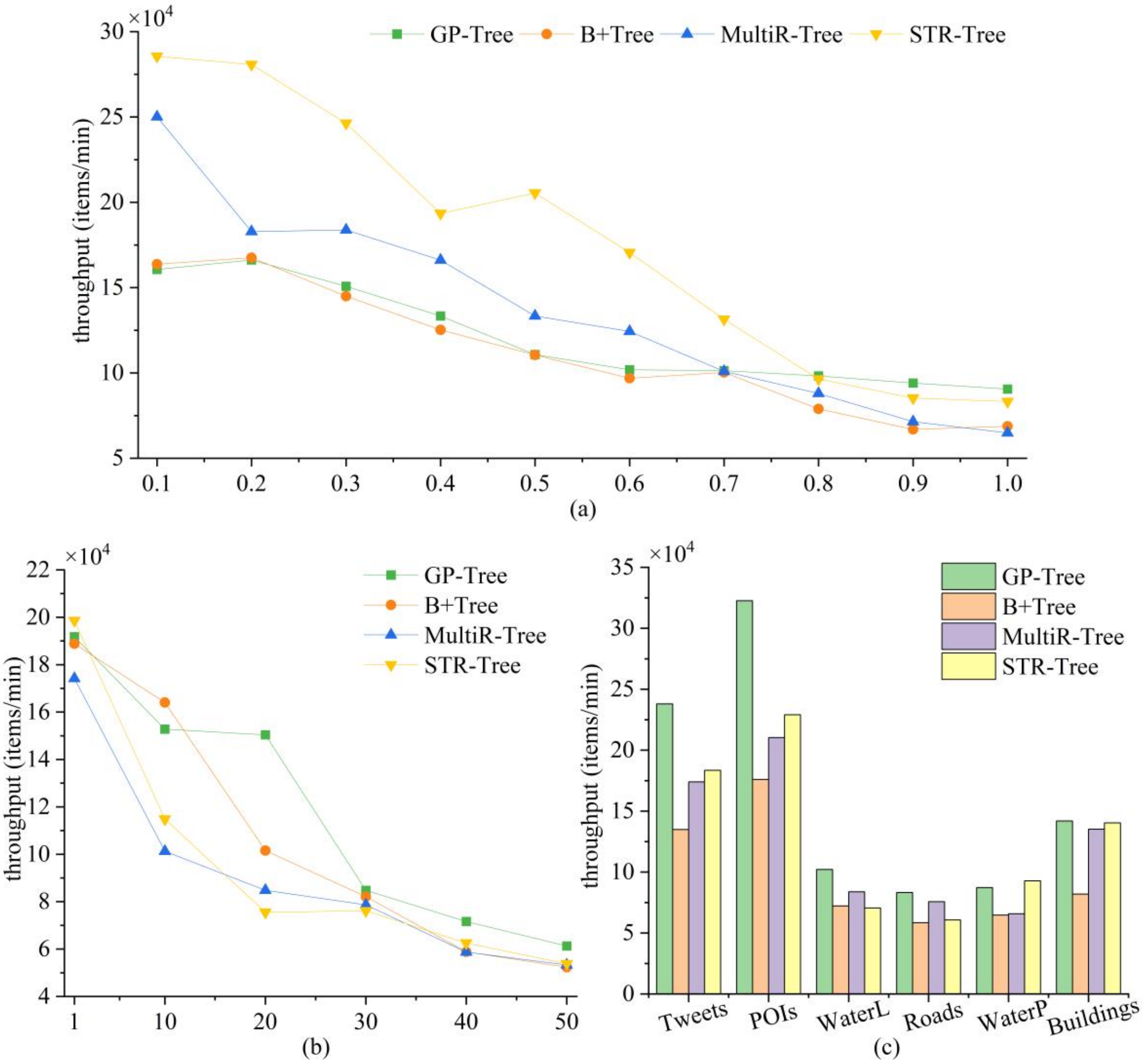

Figure 10. Comparison of different spatial indexes in kNN queries on (a) different dataset sizes, (b) different $k$ values and (c) different datasets.

**Impact of $k$ values**. Figure 10(b) illustrates the kNN throughput on the Roads dataset as $k$ varies from 1 to 50. The throughput of all indexes decreases as $k$ increases, since larger $k$ values expand the search region and lead to more candidates to be verified. When $k$ is small (e.g., $k \leq 10$), GP-Tree achieves throughput comparable to STR-Tree and B+Tree, while clearly outperforming MultiR-Tree. As $k$ increases beyond 10, GP-Tree consistently outperforms all baseline methods, demonstrating its effectiveness in handling kNN queries with large $k$ values.

**Impact of dataset types**. GP-Tree consistently achieves the highest throughput on most datasets, as shown in Figure 10(c). On the WaterP dataset, which contains complex polygonal objects, STR-Tree slightly outperforms GP-Tree, while GP-Tree still maintains competitive performance compared with other baselines. Overall, these results indicate that GP-Tree delivers robust and efficient kNN query performance across a wide range of dataset types, especially for point and linestring datasets.

*6.2.4. Construction time and in-memory cost of indexes*

This section evaluates the spatial indexes in terms of index construction time and memory consumption.

**Index construction time**. Figure 11(a) reports the construction time of these indexes across different datasets. STR-Tree achieves the shortest construction time because it indexes only a single MBR per spatial object and employs a bulk-loading construction strategy. The Quad-Tree performs second-best because it is also a single-entry index. GP-Tree exhibits construction performance comparable to MultiR-Tree and outperforms B+Tree. During construction, GP-Tree decodes cell encodings bit-by-bit to directly locate child nodes, avoiding costly key comparisons and reducing insertion overhead. MultiR-Tree inherits the bulk-loading advantage of STR-Tree, leading to comparable efficiency. In contrast, B+Tree treats cell encodings as strings and sorts them lexicographically, requiring character-level comparisons and leading to significantly higher construction cost.

**Memory consumption**. Figure 11(b) presents the memory usage of each index. STR-Tree and Quad-Tree consume the least memory, as each spatial object is represented by only a single MBR within their index structures. MultiR-Tree requires slightly more memory since each object is decomposed into multiple MBR-tiles. GP-Tree consumes less memory than B+Tree because the prefix-tree structure shares common prefixes and avoids redundant storage, whereas B+Tree stores cell encodings as complete strings, leading to larger memory footprints. We further analyze the in-memory distribution of GP-Tree, as shown in Figure 11(c). The GHSI structure occupies less than 1% of total memory and is negligible. The prefix-tree accounts for 21%–33% of memory, as it stores only object IDs and benefits from prefix sharing. The lookup table dominates memory usage (67%–77%) because it stores both spatial objects and their adaptive cell decompositions.

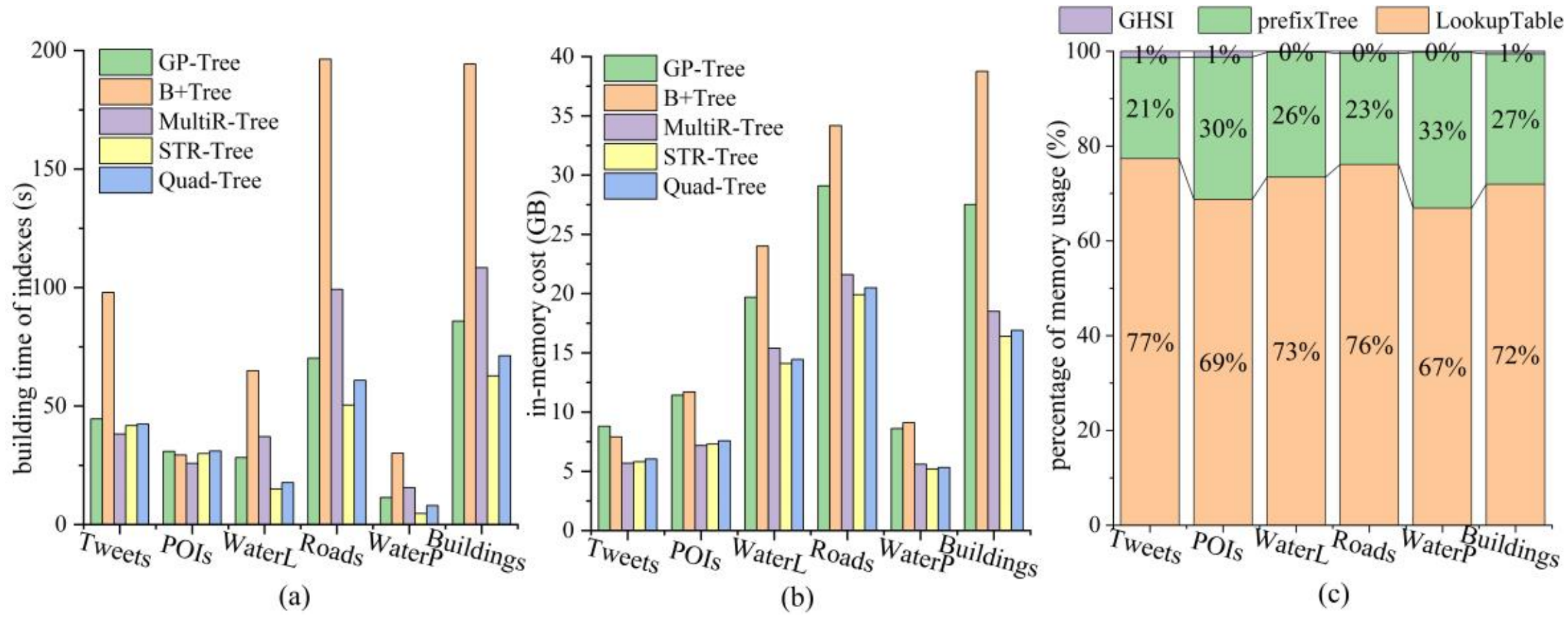

Figure 11. The comparison of different spatial indexes on (a) construction time, (b) memory cost. (c) The in-memory distribution of GP-Tree.

### *6.3. Tuning of GP-Tree*

*6.3.1. The impact of optimization strategies*

In this section, we evaluate the effectiveness of the optimization strategies proposed in Section 4.2 by comparing optimized GP-Tree (GP-Tree-Opt) and naive GP-Tree (GP-Tree-unOpt).

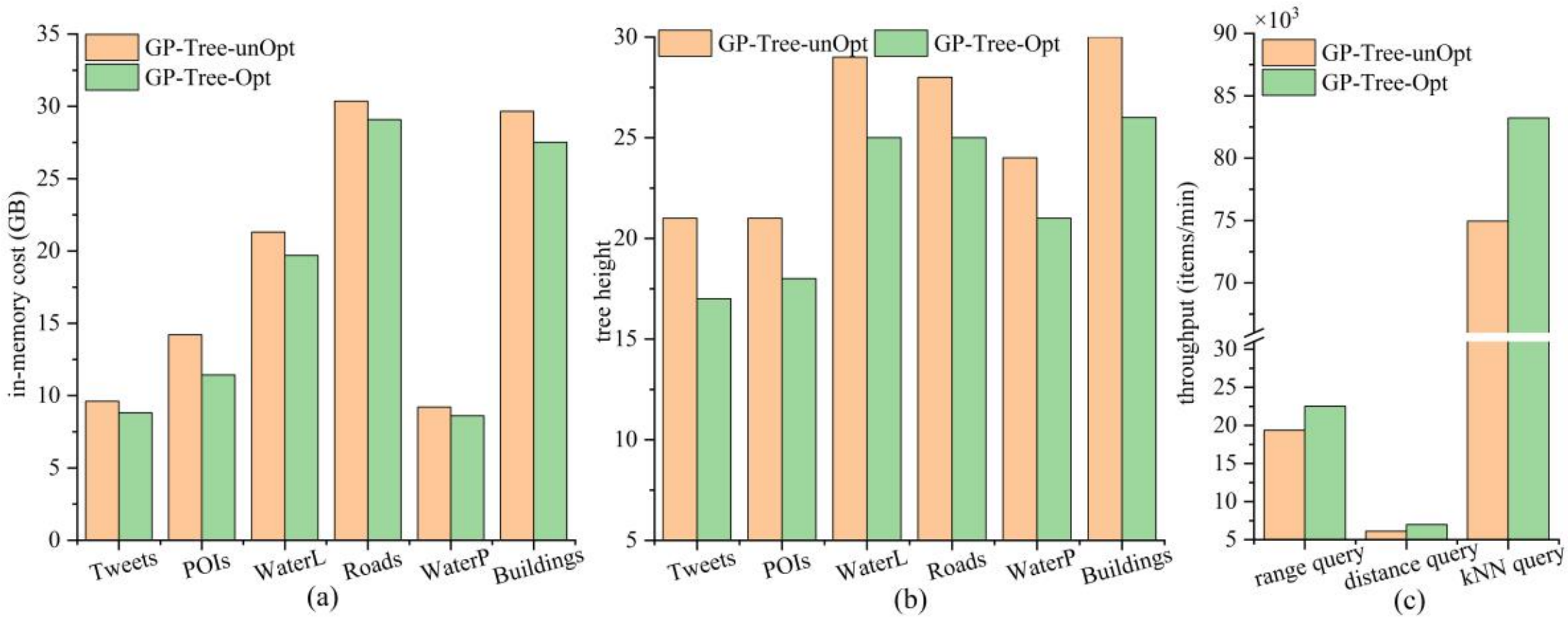

Figure 12. The impact of optimization strategies on (a) memory cost, (b) tree height and(c) efficiency of query operations.

First, the data-node optimization substantially reduces memory consumption. As shown in Figure 12(a), GP-Tree-Opt consistently requires less memory across all datasets: reductions of approximately 13.94% for point datasets, 5.87% for linestring datasets, and 6.87% for polygon datasets. Second, the pruning strategy effectively removes redundant hierarchy levels and lowers the tree height. Figure 12(b) shows that the height of GP-Tree-Opt decreases by about 16.67%, 12.25%, and 12.92% on point, linestring, and polygon datasets, respectively. Finally, the optimization strategies thereby improve query performance. As illustrated in Figure 12(c), compared with GP-Tree-unOpt, GP-Tree-Opt achieves performance gains of 16.23% for range queries, 14.38% for distance queries, and 11.03% for kNN queries. These improvements stem from two factors: (1) data-node optimization stores objects only in leaf nodes, reducing decision overhead in intermediate nodes; and (2) pruning decreases tree height, shortens search paths, and thus enhances retrieval efficiency.

*6.3.2. The impact of SEG*

This section evaluates the impact of different $SEG$ values on GP-Tree's performance. Figure 13(a) illustrates the impact of different $SEG$ values on the filtering ability of GP-Tree. As $SEG$ increases, both the FHR and THR decrease, while the proportion of UCR increases significantly. This is because a larger $SEG$ results in larger approximated grid cells (i.e., fewer grid cells per spatial object), thereby reducing the filtering effectiveness of GP-Tree.

Figure 13(b) reports the filtering time, refinement time, and total query time under different $SEG$ values for range queries. With increasing $SEG$, the number of indexed grid cells decreases, resulting in lower filtering time. However, the reduced filtering capability leads to more candidates entering the refinement phase, substantially increasing refinement time. The total query time first decreases and then increases, reflecting the trade-off between lower filtering cost and higher refinement cost; the optimal balance occurs at $SEG = 20$.

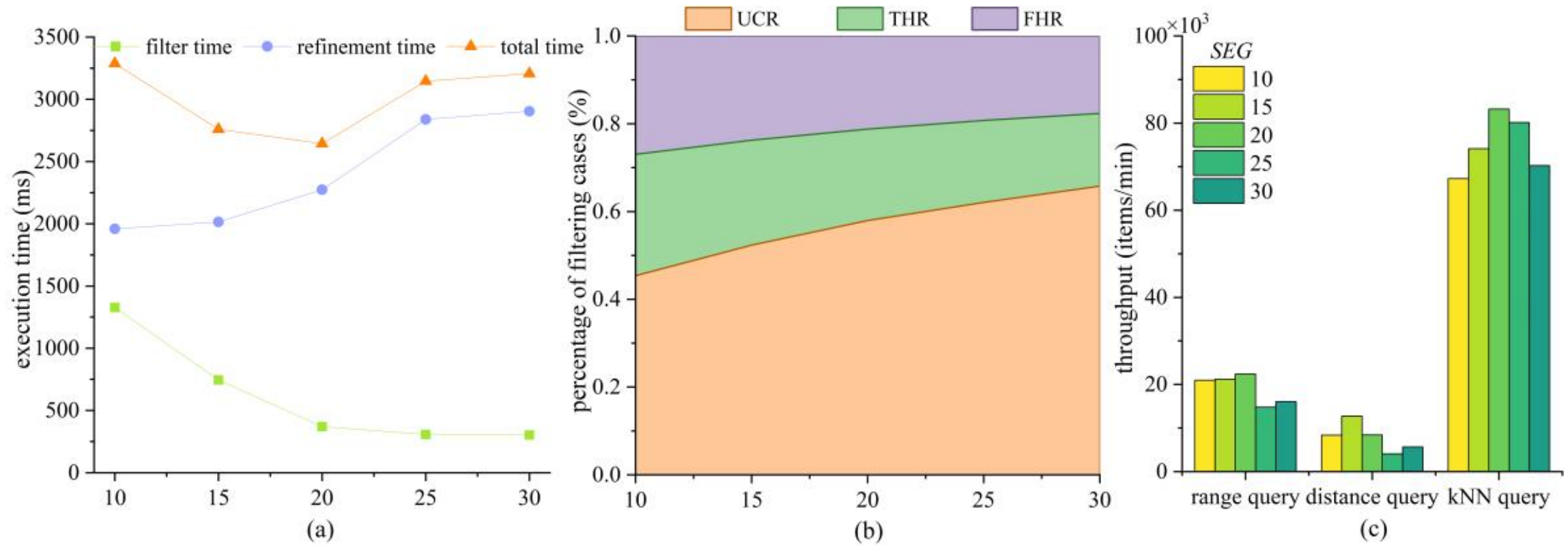

Figure 13. The impact of different *SEG* values on (a) execution times, (b) the filtering ability, and (c) different query operations of GP-Tree.

Finally, we evaluate the impact of different $SEG$ values on various spatial query operations, as shown in Figure 13(c). The throughput of all three query types increases and then decreases as $SEG$ values rise, with peak performance achieved at $SEG = 20$.

Consequently, all experiments in this chapter adopt $SEG = 20$ to ensure consistently optimal efficiency across query operations.

## 7. Conclusions and future work

This paper presents the GP-Tree to enhance the efficiency of spatial queries. GP-Tree utilizes a prefix tree to index fine-grained cell-based approximations of spatial objects, improving both the filtering capability and retrieval efficiency. We further propose two optimization strategies for GP-Tree. i.e., pruning and node optimization strategies, which reduce the tree height, shorten the search path, and lower memory overhead, thereby delivering additional performance gains. GP-Tree exhibits strong scalability and supports a wide range of spatial query operations, including range, distance, and kNN queries. Experiments on real-world datasets validate the efficiency of GP-Tree, showing that it outperforms all other spatial indexes in various query operations under our experimental setup.

Future work will proceed along two directions. First, the construction efficiency of GP-Tree is constrained by the need to index multiple approximation cells generated from each spatial object. To address this, we plan to investigate a bulk-loading strategy that accelerates index construction. Second, GP-Tree shows strong potential for extension to spatiotemporal data such as trajectories, and we aim to generalize its design to support these more complex data types.